\documentstyle[12pt]{article}
\begin{document}
\title{THE SCALED UNIVERSE}
\author{B.G. Sidharth$^*$\\
B.M. Birla Science Centre, Hyderabad 500 063 (India)}
\date{}
\maketitle
\footnotetext{$^*$E-mail:birlasc@hd1.vsnl.net.in}
\begin{abstract}
It is shown that the mysterious quantum prescription of microphysics has
analogues at the scale of stars, galaxies and superclusters, the common
feature in all these cases being Brownian type fractality. These
considerations are shown to lead to pleasingly meaningful results in agreement
with observed data.
\end{abstract}
\section{Introduction}
It has been argued by several authors\cite{r1,r2,r3,r4,r5,r6}
that it is a Brownian process that underlies quantum behaviour
and a fractal dimension of two for quantum paths. On the other hand the fractal
nature of the macro universe has been noticed over the past several years
\cite{r7,r8,r9,r10}. Indeed, this is obvious-- matter
is concentrated in atoms, stars and planets, galaxies, clusters and so on
\cite{r11} and not spread uniformly. Indeed, uniformity is dependant on the
scale of observation or resolution. Not just that:
the mysterious curiosity of a "cosmic"
quantization has also been noticed\cite{r3,r12,r13,r14}.
We will now show that the underlying connection between quantum type
phenomena at different scales has a Brownian underpinning: there exists what
may be called, a "Scaled Quantum Mechanics".
\section{Scaling}
We first observe that in Brownian motion we have\cite{r15}
\begin{equation}
x \sim \Delta x \sqrt{n}\label{e1}
\end{equation}
where, for example $\Delta x$ is the typical length of a step, $n$ is the
number of steps and $x$ is the distance covered.\\
We next observe that the following relations hold:
\begin{equation}
R \approx l_1 \sqrt{N_1}\label{e2}
\end{equation}
\begin{equation}
R \approx l_2 \sqrt{N_2}\label{e3}
\end{equation}
\begin{equation}
l_2 \approx l_3 \sqrt{N_3}\label{e4}
\end{equation}
\begin{equation}
R \sim l\sqrt{N}\label{e5}
\end{equation}
where $N_1 \sim 10^6$ is the number of superclusters in the universe,
$l_1 \sim 10^{25}cms$ is a typical supercluster size $N_2 \sim 10^{11}$ is the
number of galaxies in the universe and $l_2  \sim 10^{23}cms$ is the typical size
of a galaxy, $l_3 \sim 1$ light year is a typical distance between stars and
$N_3 \sim 10^{11}$ is the number of stars in a galaxy, $R$ being the radius
of the universe $\sim 10^{28} cms, N \sim 10^{80}$ is the number of elementary
particles, typically pions in the universe and $l$ is the pion Compton
wavelength.\\
Equations (\ref{e5}) and (\ref{e1}) have been compared and it has been argued
\cite{r5} that this is symptomatic of quantum behaviour. Here, the Compton
wavelength is a length scale within which we can find the corresponding
mass. In this same spirit
and in the light of the comments in Section 1, we can expect that equations
(\ref{e2}) to (\ref{e4}) would also lead to a quantum type behaviour though,
not at the micro scale represented by the pion (as in equation (\ref{e5}))
but rather at a suitable higher scale.\\
Infact as we will now show, this is indeed the case with a scaled Planck
constant given by
\begin{equation}
h_1 \sim 10^{93}\label{e6}
\end{equation}
for super clusters;
\begin{equation}
h_2 \sim 10^{74}\label{e7}
\end{equation}
for galaxies and
\begin{equation}
h_3 \sim 10^{54}\label{e8}
\end{equation}
for stars.\\
Let us start with equation (\ref{e5}). It is quite remarkable that (\ref{e5}),
(and a corresponding equation with the radius of the universe replaced by its
age and the Compton wavelength replaced by its Compton time) can be deduced in
a cosmological scheme in which elementary particles, typically pions are
fluctuationally produced out of a background Zero Point Field\cite{r16,r17,r18,r19}.
This scheme is consistent with astrophysical observations
and also deduces from theory the various large number relations which were
hitherto considered to be magical coincidences. Further, we have,
\begin{equation}
M = Nm,\label{e9}
\end{equation}
where $M$ is the mass of the universe and $m$ the pion mass and $N$ is defined
in (\ref{e5}). From (\ref{e9}) and (\ref{e5}) we can deduce,
\begin{equation}
\left(\frac{R}{l}\right)^2 \approx \frac{M}{m}\label{e10}
\end{equation}
From (\ref{e10}), we can easily deduce that there is the scaled Planck constant
$h_1$ given in (\ref{e6}), such that,
\begin{equation}
R = \frac{h_1}{Mc}\label{e11}
\end{equation}
Equation (\ref{e11}) shows that with this scaled constant $h_1$, the radius of
the universe turns out to be the counterpart of the Compton wavelength. Earlier
it was argued\cite{r20,r21} that an electron could be modelled as a Kerr-Newman
Black Hole with radius given by the Compton wavelength. It is interesting
to note that this is also true for the universe itself with the scaled
Compton wavelength: Infact in the case of the electron, the spin was given
by
\begin{equation}
S_K = \int \epsilon_{klm} x^l T^{m0} d^3 x = \frac{h}{2}\label{e12}
\end{equation}
where the domain of integration was a sphere of radius given by the
Compton wavelength\cite{r20,r22}. If this is carried over to the case of the
universe, with radius given by (\ref{e5}) or (\ref{e11}) and mass as in (\ref{e9}) we get
from (\ref{e12})
\begin{equation}
S_U = N^{3/2} h \approx h_1\label{e13}
\end{equation}
where $h_1$ is as in (\ref{e6}) and $S_U$ denotes the counterpart of electron spin.\\
Infact the origin of $h_1$ is in
(\ref{e13}): From this point of view, (\ref{e6}) is not
mysterious. In this case $h_1$ turns out to be the spin of the universe
itself in broad agreement with Godel's spin value for Einstein's equations
\cite{r23}. Incidentally this is also in agreement with the Kerr limit of the
spin of the rotating Black Hole of mass given by (\ref{e9}). Further as pointed out
by Kogut and others, the angular momentum of the universe given in
(\ref{e13}) is compatible with a rotation from the cosmic background
radiation anisotropy\cite{r24}. Finally it is also close to the observed rotation
as deduced from anisotropy of cosmic electromagnetic radiation as reported
by Nodland and Ralston and others\cite{r25}.\\
We next use (\ref{e3}), and the well known fact that $G \frac{m_G}{l_2} \sim
v^2$ \cite{r26}, along with the relation,
\begin{equation}
m_G v l_2 = h_2,\label{e14}
\end{equation}
which is the analogue of quantized angular momenta.
It immediately follows that $h^2_2 = G^2 m^3_G l \sim 10^{74}$, which
gives equation
(\ref{e7}). Further from (\ref{e14}) it follows that
$$v \sim 10^7 cms,$$
which is consistent with the quantized large scale velocities that have been observed
\cite{r13}.\\
Similarly taking the cue from (\ref{e4}), if in conjunction with the well known
Kepler type equation viz.,
$$G\frac{m_S}{r} \sim V^2,$$
where $m_S$ is the mass of the sun, we use the counterpart of equation
(\ref{e14}) for the sun,
can get the relation (\ref{e8}), and then the planetary angular momentum,
$\sim nh_3$ gives correctly what may be called the BodeTitius type relation for the
planetary distances. This was noted by Nottale, Agnese, Festa, Laskar, Carneiro
and others\cite{r3,r14,r12} and so will not be elaborated here. The
above considerations not only provide a rationale for this behaviour, but
also show how this fits into a more generalized scaling principle.\\
\section{Discussion}
(i) We have given a rationale for quantum like behaviour at large scales,
expressed by equations (\ref{e6}), (\ref{e7}) and (\ref{e8}). These express
scales at the level of superclusters (and the universe), galaxies and stars.
It may be mentioned that a hierarchical structure in the context of
the now defunct strong gravity
was considered by Caldirola and coworkers\cite{r27}.\\
(ii) We saw in Section 2 that the universe shows up as a black hole. Indeed
this has been argued from an alternative view point (cf.ref.\cite{r18}). Moreover,
this is symptomatic of a holistic or Machian behaviour (cf.ref.\cite{r3}). This
infact has been the purport of earlier considerations (cf.refs.\cite{r5} and
\cite{r19}, for example).\\
Moreover, this universal black hole could just be a scale expressing the upper
limit of our capability to observe. In the spirit of reference \cite{r17}, there
could be several such black holes or parallel universes.\\
In this context, we may refer to the fact that in earlier work (cf. for example
refs.\cite{r17} and \cite{r5}), a background zero point field (ZPF) or
Prigogine's quantum vacuum was considered, our of which elementary particles,
typically pions were fluctuationally created - the energy of the ZPF within
the pion Compton wavelength being the rest energy of the pion.\\
We could turn the perspective around and consider the creation of the universal
black hole instead, at the scaled "Compton" wavelength of the universe, as
given by equation (\ref{e11}). It would then appear that the structure
formation of the universe would be due to, as Mandelbrot pointed out, a
curdling process (cf.ref.\cite{r11} and \cite{r28}).\\
Indeed, we are prisoners of perspective: the following analogy would clarify.
Let us consider the drawing on (two dimensional) paper of a (three dimensional)
match box or rectangular parallelopiped. There are two rectangles - the outside
rectangle and the inside one. Depending on which of these we start with, the
match box would be either going inwards or coming outwards, two totally
different possibilities.\\
(iii) Given equations like (\ref{e6}), (\ref{e11}) and subsequent considerations
of Section 2, it would be natural to expect the universe to be a "wave
packet", though not in the spirit of Hawking\cite{r29}. We can see that this
is indeed so. Infact for a Gaussian wave packet\cite{r30}, we have,
\begin{equation}
R \approx \frac{\sigma}{\sqrt{2}} \left( 1+ \frac{h^2_1 T^2}{\sigma^4 M^2}\right)^{1/2}
\left( \approx \frac{1}{\sqrt{2}} \frac{h_1 T}{\sigma M}\right)\label{e15}
\end{equation}
where, now, $R$ and $T$ denote the radius and age of the universe (at a given
time), $M$ its mass and $\sigma \sim R$ is the spread of the wave packet.\\
Remembering that $R \approx cT$, (\ref{e15}) actually reduces to equation
(\ref{e11})! The width of the wave packet is the "Compton" length. Differentiation
of (\ref{e15}) gives,
\begin{equation}
\dot R \approx \frac{h_1}{\sigma M} \approx \frac{h_1}{M\sigma^2} \cdot R \equiv
HR\label{e16}
\end{equation}
Equation (\ref{e16}) resembles Hubble's law. We can show that this is indeed
so: Using (\ref{e9}), and (\ref{e13}), it follows from (\ref{e16}) that,
\begin{equation}
H \approx \frac{c}{\sqrt{N}l} \approx \frac{Gm^3c}{h^2}\label{e17}
\end{equation}
where the last equality has been deduced previously (cf.refs.\cite{r16,r17}) and
can be easily verified.
Not only does (\ref{e17}) give the correct value of the Hubble constant, but it
is also Weinberg's "mysterious" empirical relation (cf.ref.\cite{r26}), giving
the pion mass in terms of the Hubble constant or vice versa.\\
Interestingly, from (\ref{e15}), using again $R \approx cT$, we can deduce that
\begin{equation}
Mc^2 \cdot T \approx h_1\label{e18}
\end{equation}
Equation (\ref{e18}) and (\ref{e11}) are the analogues of Heisenberg's
Uncertainity Principle.

\end{document}